\documentclass[twocolumn,english,aps,prb,superscriptaddress]{revtex4-1}
\usepackage{ifpdf}
\usepackage{graphicx}
\usepackage{epstopdf}
\usepackage{grffile}
\usepackage[latin9]{inputenc}
\usepackage{color}
\usepackage{babel}
\PassOptionsToPackage{normalem}{ulem}
\usepackage{ulem}
\usepackage[unicode=true, bookmarks=false, breaklinks=false,pdfborder={0 0 1},colorlinks=false] {hyperref}
\usepackage{breakurl}
\usepackage{amsmath}
\begin{document}

\title{In-plane antiferromagnetic moments in axion topological insulator candidate EuIn$_2$As$_2$}

\author{Yang Zhang}
\email{The authors contribute to equal}
\affiliation{School of Physics, Sun Yat-Sen University, Guangzhou, Guangdong 510275, China}
\author{Ke Deng}
\email{The authors contribute to equal}
\affiliation{Shenzhen Institute for Quantum Science and Engineering (SIQSE) and Department of Physics, Southern University of Science and Technology, Shenzhen 518055, China}
\author{Xiao Zhang}
\affiliation{School of Physics, Sun Yat-Sen University, Guangzhou, Guangdong 510275, China}
\author{Meng Wang}
\affiliation{School of Physics, Sun Yat-Sen University, Guangzhou, Guangdong 510275, China}
\author{Yuan Wang}
\affiliation{Shenzhen Institute for Quantum Science and Engineering (SIQSE) and Department of Physics, Southern University of Science and Technology, Shenzhen 518055, China}
\author{Cai Liu}
\affiliation{Shenzhen Institute for Quantum Science and Engineering (SIQSE) and Department of Physics, Southern University of Science and Technology, Shenzhen 518055, China}
\author{Jia-Wei Mei}
\affiliation{Shenzhen Institute for Quantum Science and Engineering (SIQSE) and Department of Physics, Southern University of Science and Technology, Shenzhen 518055, China}
\author{Shiv Kumar}
\affiliation{Hiroshima Synchrotron Radiation Center, Hiroshima University, Higashihiroshima, Hiroshima 739-0046, Japan}
\author{Eike F. Schwier}
\affiliation{Hiroshima Synchrotron Radiation Center, Hiroshima University, Higashihiroshima, Hiroshima 739-0046, Japan}
\author{Kenya Shimada}
\affiliation{Hiroshima Synchrotron Radiation Center, Hiroshima University, Higashihiroshima, Hiroshima 739-0046, Japan}
\author{Chaoyu Chen}
\email{Corresponding author: chency@sustech.edu.cn}
\affiliation{Shenzhen Institute for Quantum Science and Engineering (SIQSE) and Department of Physics, Southern University of Science and Technology, Shenzhen 518055, China}
\author{Bing Shen}
\email{Corresponding author: shenbingdy@mail.sysu.edu}
\affiliation{School of Physics, Sun Yat-Sen University, Guangzhou, Guangdong 510275, China}

\begin{abstract}
Topological insulator with antiferromagnetic order can serve as an ideal platform for the realization of axion electrodynamics. In this paper, we report a systematic study of the axion topological insulator candidate EuIn$_2$As$_2$. A linear energy dispersion across the Fermi level confirms the existence of the proposed hole-type Fermi pocket. Spin-flop transitions occur with magnetic fields applied within the $ab$-plane while are absent for fields parallel to the $c$-axis. Anisotropic magnetic phase diagrams are observed and the orientation of the ground magnetic moment is found to be within the $ab$-plane. The magnetoresistivity for EuIn$_2$As$_2$ behaves non-monotonic as a function of field strength. It exhibits angular dependent evolving due to field-driven and temperature-driven magnetic states. These results indicate that the magnetic states of EuIn$_2$As$_2$ strongly affect the transport properties as well as the topological nature.
\end{abstract}
\pacs{}
\date{\today}
\maketitle

\section{Introduction}
Axion topological insulator (ATI) is an exotic material which can host hypothetic quasiparticles axions within the Standard Model of particle physics \cite{X.-L. Qi, Rundong Li}. Axion electrodynamics emerge on the surface of topological insulators with magnetic order. It can be described by the formula
\begin{equation}
S_\theta=\frac{\theta e^2}{4 \pi^2}\int dtd^3 \mathbf{x}\mathbf{E} \cdot \mathbf{B}
\end{equation}
where $\mathbf{E}$ and $\mathbf{B}$ are electromagnetic fields and the axion angle $\theta$=$\pi$ gives rise to topological magnetoelectric (TME) effect \cite{X.-L. Qi,F. Wilczek}. The ATI is realizable in various artificial structures such as sandwich heterostructures but still unreachable in the intrinsic materials up to now \cite{M. Mogi, Di Xiao}. Recently theoretical calculations predicted that axion topological states were able to exist in the intrinsic materials with antiferromagnetic (AF) order spontaneously breaking the time-reversal symmetry (TRS)\cite{Dongqin Zhang,Sugata Chowdhury,Alexander Zeugner,M. M. Otrokov,Yuanfeng Xu,Jiaheng Li}.

The van der Waals (vdW) Mn-Bi-Te family of materials were proposed magnetic topological insulators with the intrinsic $A$-type AF order, which could host quantum anomalous Hall effect (QAHE) with a non-zero Chern number \cite{Jiaheng Li}. Recent neutron experiments observed this $A$-type magnetic order in MnBi$_2$Te$_4$ agreeable with the prediction of ATI\cite{J.-Q. Yan}. And in the ultra thin flakes or films of MnBi$_2$Te$_4$, a moderate magnetic field could drive the system into a Chern insulating state exhibiting zero longitudinal resistance and quantized Hall resistance $h/e^2$ (QAHE) as well as favoring the former prediction \cite{Yujun Deng, Jun Ge, Chang Liu}. Nonetheless, subsequent Angle-resolved Photoelectron Spectroscopy (ARPES) results indicated that the Dirac surface states were gapless crossing the projected bulk band gap \cite{R. C. Vidal,Bo Chen,Yan Gong, Hao Li,YL Chen,Kaminski,Hong Ding,Yu-Jie Hao,Chaowei Hu}. Moreover, the neutron scattering experiments revealed that the short range magnetic order existed in a wide temperature region companying with strong magnetic fluctuation in MnBi$_2$Te$_4$ \cite{J.-Q. Yan1}. These results indicate that the magnetic surface configuration and the topological states in MnBi$_2$Te$_4$ should be treated carefully.

Different from layered vdW MnBi$_2$Te$_4$, EuIn$_2$As$_2$ exhibits a three-dimensional structure and crystallizes in the $P6(3)/mmc$ space group (in Fig. 1(a))\cite{Andrea M. Goforth}. It is an AF topological insulator candidate \cite{Hong Ding, Xin Gui} with tunable high-order topological insulating states (HOTIS) and axion topological insulating states (ATIS) \cite{Yuanfeng Xu}. Particularly, the non-trivial topological states are strongly influenced by the detailed magnetization according to the theoretical calculations\cite{Yuanfeng Xu}. Although the ground state AF order is along the $c$-axis for EuIn$_2$As$_2$, manipulating the direction of magnetic moments allows for archiving different topological states: in-plane magnetic moments result in an ATIS  with gapped surface states while the out-of-plane magnetic moments lead to a HOTIS with hinge state on the domain wall of two gapped surfaces \cite{Yuanfeng Xu}. The investigation of the detailed magnetization for EuIn$_2$As$_2$ is needed and necessary to gain the insight of its topological nature. In this paper, we systemically study EuIn$_2$As$_2$ by combining magneto-transport measurements and angle-resolved photoemission spectroscopy measurements . A spin-flop transition is observed by applying small magnetic fields within the $ab$-plane. Anisotropic magnetic phase diagrams reveal in-plane AF moment agreeable with the ATI prediction.  The field and temperature driven magnetic states dominate the transport properties in low temperature region.  The observed hole-type Fermi surface made up of a linear dispersing band indicates that pristine EuIn$_2$As$_2$ is a metal instead of insulator. Further tuning chemical potential such as doping or gating the system may realize ATIS in this intrinsic material.

\section{Experimental Detail}

\begin{figure}
  \centering
  \includegraphics[width=3.5 in]{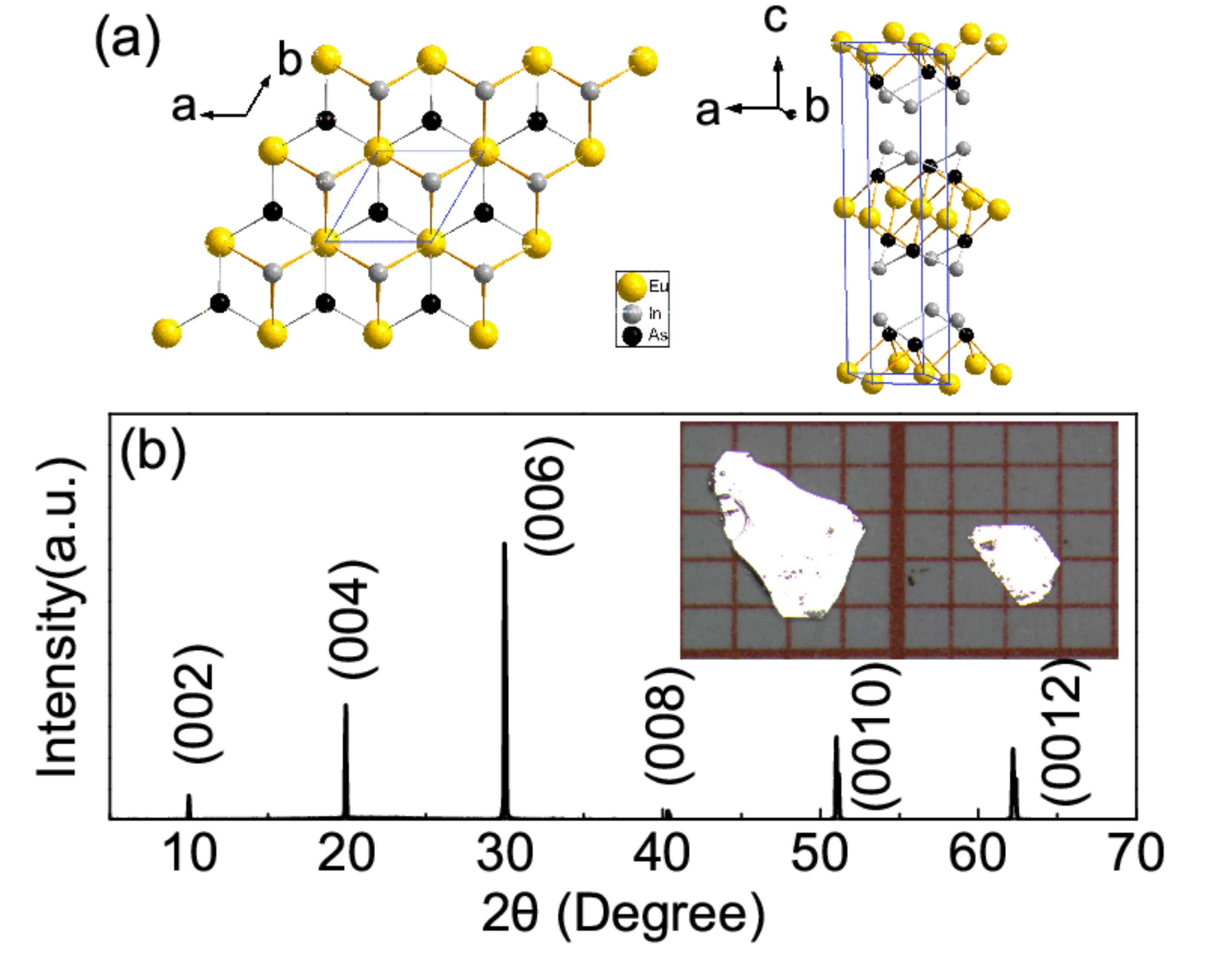}
  \caption{(a) Crystal structure of EuIn$_2$As$_2$. (b) X-ray diffraction pattern of a single crystaline EuIn$_2$As$_2$. High crystallinity is indicated by the set of (00l) peaks. Inset: picture of typical single crystals of EuIn$_2$As$_2$.
 }
  \label{fig:Fig1}
\end{figure}

\begin{table}
\centering
\caption{The  crystal data and structure refinement for EuIn$_2$As$_2$ at 150 K.}
\label{my-label}
\begin{tabular}{c c}
\hline
\hline
Empirical formula   & EuIn$_2$As$_2$\\

Formula weight   & 531.44 \\

Crystal system  & hexagonal\\

Space group & P63/mmc   \\

unit cell dimensions  & a=4.2068(5)${\textrm{\AA}}$ \\
                      & b=4.2068(5)${\textrm{\AA}}$ \\
                      & c=17.849(3)${\textrm{\AA}}$  \\
$\alpha$ & 90$^o$ \\
$\beta$ & 90$^o$ \\
$\gamma$ & 120$^o$ \\
Volume & 273.55(8) ${\textrm{\AA}}^3 $  \\

Z & 2   \\

$\rho$(calc)   & 6.452 g/cm$^3$  \\

$\mu$  & 31.511  mm$^{-1}$\\

F(000)  & 454.0   \\

Radiation & MoK $\alpha$ ($\lambda$= 0.71073)    \\

2$\theta$ range for data collection  & 9.136$^o$ to 69.908$^o$    \\

Index ranges  & 6 $\leq$ h $\leq$ 6, \\
              &-5 $\leq$ k $\leq$ 3, \\
              & -20 $\leq$ l $\leq$ 28     \\

Independent reflections  & 271 [R$_{int}$ = 0.0462, \\
                         &R$_{\sigma}$ = 0.0443]     \\
Data/restraints/parameters & 271/0/10 \\
Goodness-of-fit on F$^2$  & 1.023    \\

Final R indexes [I $\geq$ 2$\sigma$ (I)]  & R$_1$ = 0.0315,\\
                                          & $\omega$R$_2$ = 0.0741     \\

Final R indexes [all data]   & R$_1$ = 0.0434,\\
                             & $\omega$R$_2$ = 0.0784     \\

Largest diff. peak/hole / e ${\textrm{\AA}}^3 $ &1.63/-1.87      \\
\hline
\hline
\end{tabular}
\end{table}

EuIn$_2$As$_2$ single crystals were grown via the self-flux method. Eu, In, and As were mixed stoichiometrically and sealed inside an evacuated quartz tube. The mixture was heated to 1000 $^o$C, slowly cooled to 700 $^o$C, and finally decanted by a centrifuge.  Planar single crystals with the dimensions of 2$\times$3$\times$0.3  mm$^3$ were harvested as shown in the inset of Fig.1 (b).  Geometric structure and elemental composition were confirmed by X-ray diffraction (XRD) and Wavelength-dispersive spectroscopy (WDS). Sharp peaks in the XRD confirm high crystalline quality of the samples (Fig.~\ref{fig:Fig1} (b)).

Detailed structural information was summarized in Table 1. Rietveld refinements were carried out with the SHELXL software. ARPES measurements were performed at Beamline 1 of Hiroshima Synchrotron Radiation Center (HSRC), Hiroshima University, Japan, with a VG Scienta R4000 electron analyzer \cite{Iwasawa}. The photon energy was fixed at 30 eV. The beam was linearly polarized with its polarization lying in the measurement and incidence plane. The energy and angular resolutions were set at 25 meV and 0.2$^o$, respectively. Samples were cleaved at 30 K and measured at 20 K, both at vacuum better than 2$\times$10$^{-11}$ mbar. For the magneto transport measurements rectangular samples were selected to fabricate into devices to avoid influence of geometry on the results (Fig.2 (a)). Electronic contacts were added via patterned mask with photolithography subsequent growth of Ti/Au(10nm/100nm) layers.  Electrical transport measurements were performed in Quantum design Physical Properties Measurement System (QD PPMS Dynacool). Direct current (DC) magnetization and alternating current (AC) susceptibility measurements were performed in Quantum Design Magnetic Property Measuring System (MPMS).

\begin{figure}
  \centering
  \includegraphics[width=3.1 in]{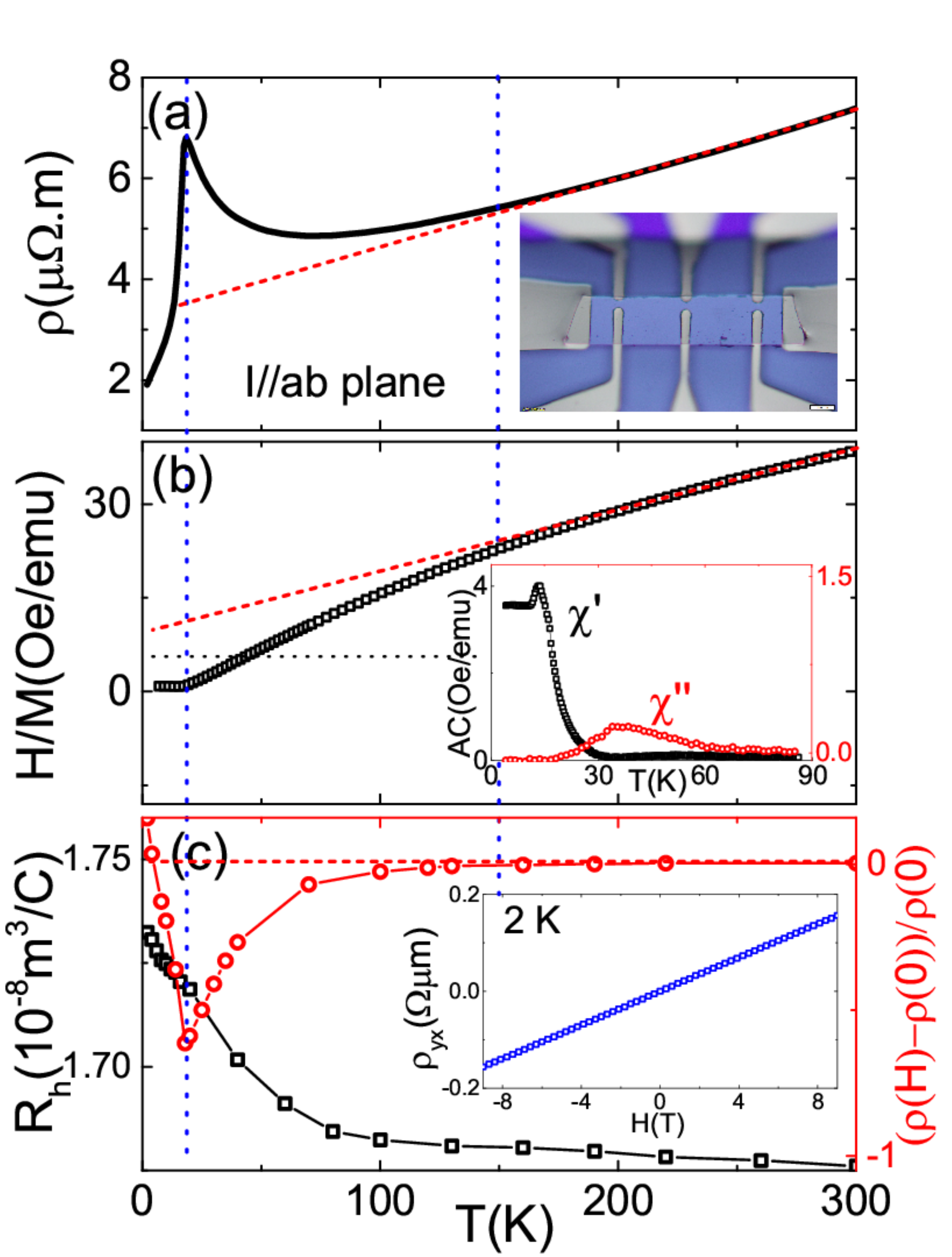}
  \caption{(a) Temperature dependent resistivity of EuIn$_2$As$_2$ within the $ab$-plane. Above 150 K the curve is  quasi linear. The red dashed line is used to estimate the deviation from the linear behaviour below 150 K. Inset: Picture of EuIn$_2$As$_2$ with patterned contacts used for the measurement. (b) Temperature dependent DC magnetization of EuIn$_2$As$_2$ with magnetic field of $H$ = 1 T applied within the $ab$-plane. Deviation from the Curie-Weiss behaviour (marked by the red dashed line) is observed below 150 K. Inset: Temperature dependent AC susceptibility measured  with an amplitude of 1 Oe and a frequency of 1000 Hz. The imaginary part $\chi''$ is zero below 18 K and exhibits a broad peak around 35 K. (c) Temperature dependence of MR at 1 T with $H//c$ is plotted in red. Hall coefficient is plotted in black, respectively.  Obvious MR is observed below 150 K and a transition from negative to positive MR occurs around 18 K. The inset: Field dependent Hall resistivity $\rho_{xy}$ at 2 K exhibits a linear behavior.   }
  \label{fig:Fig2}
\end{figure}

\section{Results and discussion}
Figure 2 shows the temperature dependent resistivity $\rho$($T$) and magnetization of EuIn$_2$As$_2$. The system behaves metallic in the high temperature region (shown in Fig.1 (a)) and exhibits a peak in $\rho$($T$) at 18 K indicative of an AF transition reported previously \cite{Andrea M. Goforth}. The temperature dependent DC magnetization $M$($T$) and AC susceptibility $\chi'$($T$) shown in Fig. 2 (b) are consistent with this transition to an AF ground state. Above the AF transition temperature $T_{N}$, an obvious broad peak of the imaginary part of the AC susceptibility $\chi''$  is observed around 40 K. The related non-zero response of $\chi''$  persists  up to the temperatures far above $T_{N}$ due to Eu$^{2+}$ spin dynamics. The $M$($T$) curve as well as starts to deviate from Curie Weiss behavior below 150 K (also far above $T_{N}$ ) which is linked with the deviation from a quasi-linear temperature dependence $\rho$($T$)  shown in Fig. 1 (a). These results indicate that the strong short-range AF fluctuation state may exist in a wide temperature region extending well above $T_{N}$ inconsistent instead of the paramagnetic state. Such a scenario would explain the observed anisotropic $g$-shift and a line width broadening above $T_{N}$ revealed by Electron spin resonance (ESR) measurements\cite{P. F. S. Rosa}. Further weight is given to the proposed SR-AF phase by the temperature dependent magnetoresistivity (MR)  measurements in Fig. 2 (c). The MR (defined as MR=$\rho$($H$)/$\rho$(0 T)-1) above 150 K is negligible consisted with the paramagnetic state while it becomes obvious below 150 K and reach to the largest negative value at $T_{N}$. Below $T_{N}$ the negative MR is suppressed when EuIn$_2$As$_2$ enters the AF ground state. It changes the sign to positive at 4 K and reaches up to the largest positive value of 30\% at 2 K. This behavior was  observed in antiferromagnets crossing the metamagnetic phase transition due to appearance of various local magnetic moments\cite{A.P.Pikul}. The positive Hall coefficient shown in Fig. 2 (d) exhibits an slight upturn of less than 3\% towards lower temperatures. This small upturn may be resulted from local spin alignment changing the carrier effective mass or carrier concentration. Both carrier density and effective mass can be determined by measurements of the temperature dependence of the electronic band structure by Angle-Resolved Photoelectron Spectroscopy (ARPES). However, the variation of less than 3\% is difficult to resolve in our ARPES measurements and out of the focus of the current work.

  \begin{figure}
  \centering
  \includegraphics[width=3.5 in]{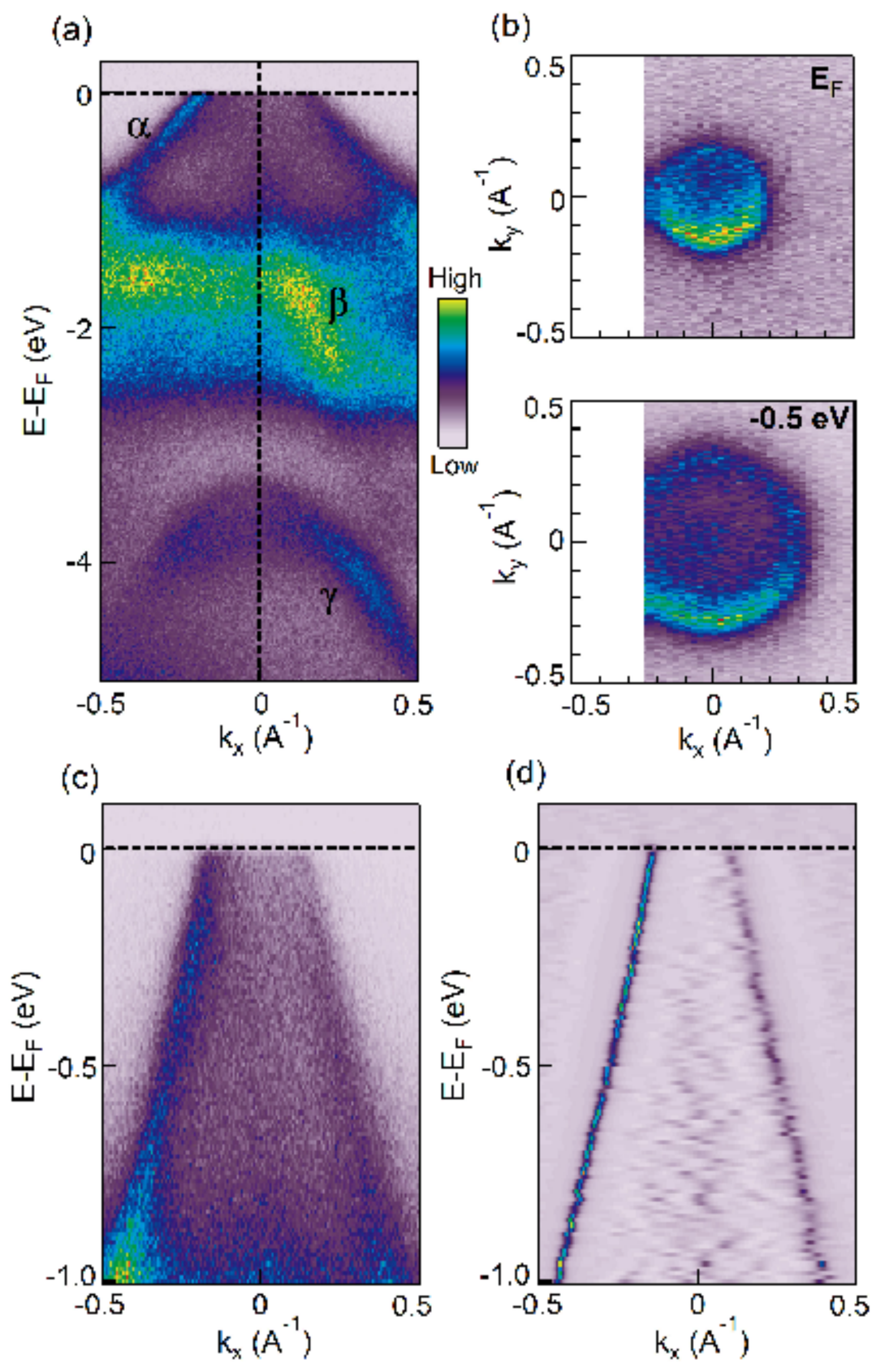}
  \caption{(a) Valence band structure along the K-G-K high symmetry direction. Valence bands are labeled from low toward higher binding energy as $\alpha$, $\beta$, and $\gamma$. The $\alpha$ state is surface derived as discussed in previous work \cite{Hong Ding}.
(b) Fermi surface map (top) and constant energy contour at 0.5 eV below the Fermi level. (c) Zoom-in of the surface state close to the Fermi level. (d) The corresponding MDC second derivative of (c) to highlight its linear dispersion. }

  \label{fig:Fig3}
\end{figure}

Figure 3 shows ARPES spectra of the electronic structure of EuIn$_2$As$_2$ measured at 20 K. Three valence bands $\alpha$, $\beta$, $\gamma$ (from low to high binding energy) near Fermi level were observed along the K-G-K high symmetry direction of the surface Brillouin zone (Fig.~\ref{fig:Fig3}(a)). The fully occupied $\beta$ and $\gamma$ band originate from bulk states, while the only band crossing the Fermi level ($\alpha$) is surface derived \cite{Hong Ding}. Fig.~\ref{fig:Fig3}(b) shows the constant energy contours of the band structure, with one at and the other one 500 meV below the Fermi level. A circular hole-like pocket formed by $\alpha$ grows in sizes with increasing binding energy agrees well with  Fig.~\ref{fig:Fig3}(a). A zoom in of Fig.~\ref{fig:Fig3}(a) enables us to scrutinize the dispersion of the surface band and fit the energy dispersion $\mathbf{E}$($\mathbf{k}$) using MDC cuts. The extracted $\mathbf{E}$($\mathbf{k}$) curve is linear with Fermi velocity of 3.7 eV$\cdot$A (5.7105 m/s) and $k_F$ =0.12 1/\AA.(Fig.~\ref{fig:Fig3}(c,d)). In addition, we note that by using the extracted $k_F$ and Fermi velocity, the Dirac point can be estimated as 300 meV above Fermi level. Further tuning the chemical potential by doping or gating electron would lift the Fermi level close to the Dirac point.

 \begin{figure}
  \centering
  \includegraphics[width=3.5 in]{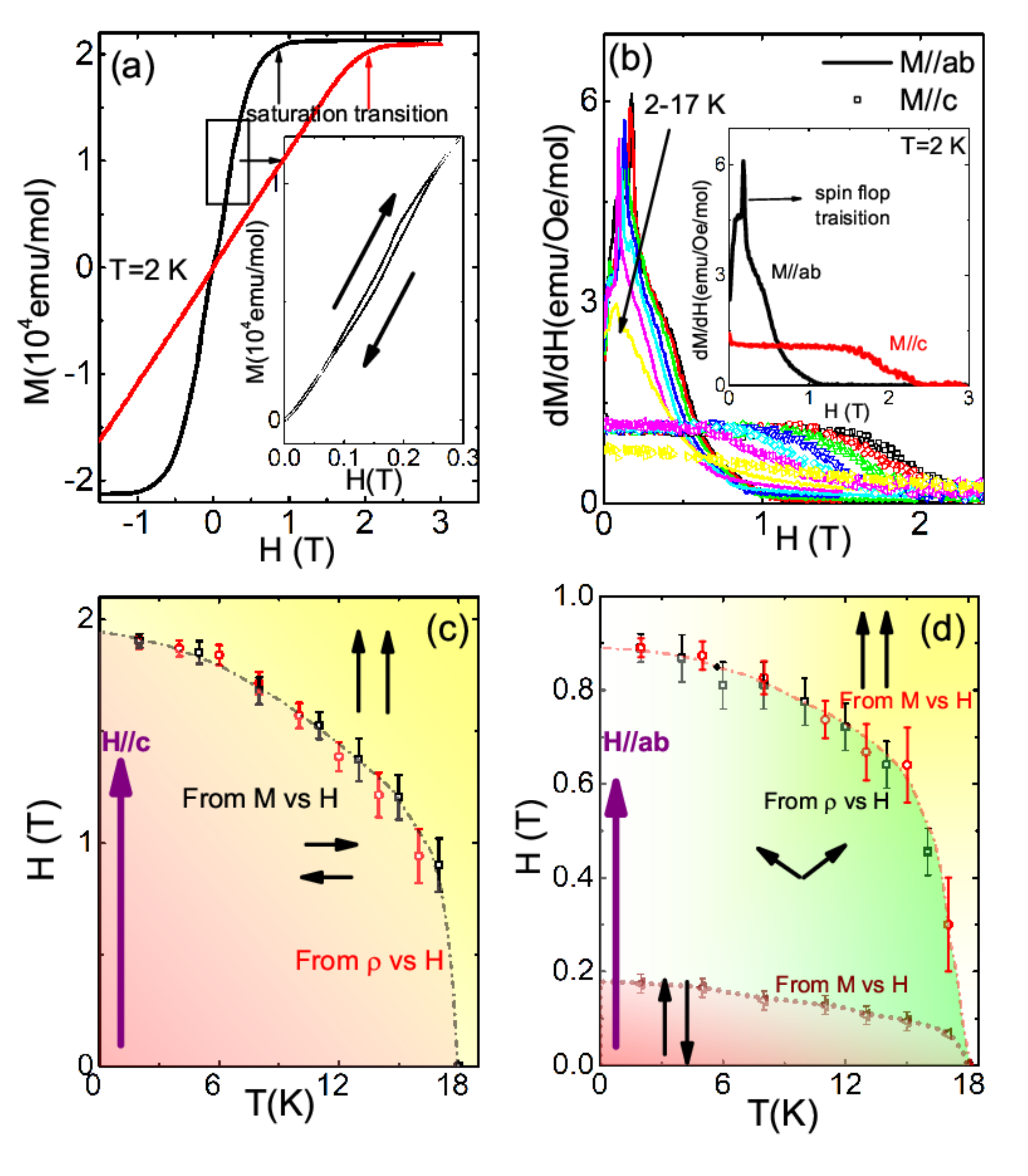}
  \caption{(a) Magnetic hysteresis loops ($MH$) for EuIn$_2$As$_2$ with $H$//$ab$ (black lines) and $H$//$c$ (red line) at 2 K. The $MH$ curves in the region marked with a rectangle is zoomed and shown in the inset. A jump of the hysteresis is observed in the MH curves, if field is applied within the $ab$-plane. (b) Derivative of $MH$ (d$M$/d$H$) at 2 K, 5 K, 8 K, 11 K, 13 K, 15 K, and 17 K with applied field within the $ab$-plane (marked with lines) and $c$-axis (marked with open symbols), respectively. Inset: d$M$/d$H(H)$ at 2 K with $H//ab$ (marked with black line) and $H//c$ (marked with red line). We can observe a sharp peak related to spin flop transition with $H$//$ab$.  (c) and (d) The phase diagram of EuIn$_2$As$_2$ with $H$//$ab$ and $H$//$c$, respectively. The purple arrow indicates the direction of applied magnetic field. The black arrows indicate the direction of Eu$^{2+}$ spins }
  \label{fig:Fig4}
\end{figure}

Theoretical calculations proposed two distinct AF ground states with different orientation of the magnetic moment but the small energy difference for EuIn$_2$As$_2$. The detailed investigation of the magnetic phase diagram is therefore vital to understand or even control the proposed non trivial topological states in this material. To this end, both DC magnetic susceptibility and magneto-transport measurements were performed in single crystals of EuIn$_2$As$_2$ with magnetic fields applied within the $ab$-plane ($H$//$ab$) and  along the $c$-axis ($H$//$c$) respectively. The field dependent magnetization $M$($H$) curves at 2 K are shown in Fig. 4(a) exhibiting anisotropic behavior. Increasing the applied magnetic field along $ab$-plane to a critical value $H_{sp}$, a small jump accompanying a magnetic hysteresis in $M$($H$) is observed, which indicates a first order magnetic transition (shown in the inset of Fig. 4(a)). This abnormal behaviour is more clearly identified by a sharp peak in the d$M$/d$H$ curves shown in Fig. 4(b).  The magnetization  saturates when further increasing the magnetic field to the critical value $H_{sa}$.  As with $H//c$, no additional magnetic transition occurs besides the  saturation (occurred at $H^\bot_{sa}$) shown in Fig. 4(a) and (b). Such anisotropic magnetic behavior can be interpreted within the picture of spin-flop transitions: in a collinear AF system,  increasing the applied field to a critical value $H_{sp}$ along, or nearly parallel to its magnetic easy axis (MEA), the AF sub-lattice magnetizations $M_1$ and $M_2$ rotate abruptly \cite{Y. Shapira}. The staggered magnetization $L=M_1-M_2$ directs perpendicular to the original MEA instead of aligning itself with the external field direction since the system energy is lower in the former case. This so called spin-flop transition is usually a first-order phase transition and can be detected by a jump in the magnetization loop or a peak in field dependent heat capacity.\cite{Y. Shapira, Hai-Feng Li}. Further increasing the applied magnetic field above $H_{sp}$ will gradually tilt the spin-flopped moments towards the direction of the external magnetic field and finally align all the spins' moments well along at $H_{sp}$. Thus, two transitions will be observed in the case of applied filed parallelled to MEA. On the other hand, with the applied field perpendicular to the MEA, the AF sublattice spins tilt gradually and finally were well aligned by external field directly at a critical field $H^\perp_{sa}$. As shown in Fig. 5(c) and (d), applying the magnetic field along different directions gives rise to the anisotropic magnetic phases diagram of EuIn$_2$As$_2$.

In a simple localized collinear AF system, the magnetization is mainly determined by AF exchange interaction,  magnetocrystalline anisotropic
energy  and Zeeman energy.  Considering the single ion case the Hamiltonian can be written as
\begin{equation}
\begin{split}
H&=-J\sum \overrightarrow{S}_i \cdot \overrightarrow{S}_j-D\sum S^2_{iz}+D\sum S^2_{jz}\\
   &-g\mu_BH_0(\sum S_{iz}+\sum S_{jz})
\end{split}
\end{equation}
where $S_i$ and $S_j$ are the spins at site $i$ and $j$ of the sublattice, $H_0$ is the applied static magnetic field along the $z$ direction, $J$ is the exchange constant of the interaction between spins $S_i$ and $S_j$ \cite{Oliveira}. The $D$-term represents the uniaxial anisotropy of a the single ion and the last term is the usual Zeeman term. With the field applied parallel to the MEA, the molecular field approximation based on a semiclassical model gives critical fields at T=0 K as:
\begin{equation}
H_{sp}(0)=(H_A(2H_E-H_A))^{0.5}
\end{equation}
\begin{equation}
H_{sa}(0)=2H_E-H_A
\end{equation}
and with  fields applied perpendicular to the MEA, only the single critical field $H^\bot_{sa}$(0) exists and can be expressed as:
 \begin{equation}
H^\bot_{sa}(0)=2H_E+H_A
\end{equation}
where the  $H_E$=$zJ/g$$\mu_B$ is the exchange field, $H_A$= $D$/$g\mu_B$ is the anisotropy field, and $z$ is the number of nearest neighbors.
By extrapolating the experimental $H_{sp}$($T$),  $H_{sa}$($T$), and $H^\bot_{sa}$($T$) curves to 0 K in Fig 4.(c) and (d), the $H_{sp}$(0) and $H_{sa}$(0) and $H^\bot_{sa}$(0) are estimated to be 0.18 T, 1.05 T, and 2.1 T respectively. Employing formulas (2) and (4), we calculate that $H_A$=0.03 T and $H_E$= 0.54 T.
According to formula (3), the calculated $H_{sp}$(0)=0.7 T is much larger than the experimental value of 0.19 T. It is reminiscent of the overestimated values in nickel oxide which was attributed to the existence of abnormal spin wave modes \cite{Machado}.  The presence of magnetic fluctuation above $T_{N}$ as discussed above, together with the mismatch of experimental and theoretical $H_{sp}$ give weight to the hypothesis that EuIn$_2$As$_2$ is in a state of fluctuating spin-waves. Such a non-standard spin wave excitation can not be understood by semiclassical picture and would explain the deviation of the extracted critical fields. Inelastic neutron scattering experiments are needed and necessary to probe the spin excitation in EuIn$_2$As$_2$.

 \begin{figure}
  \centering
  \includegraphics[width=3.5 in]{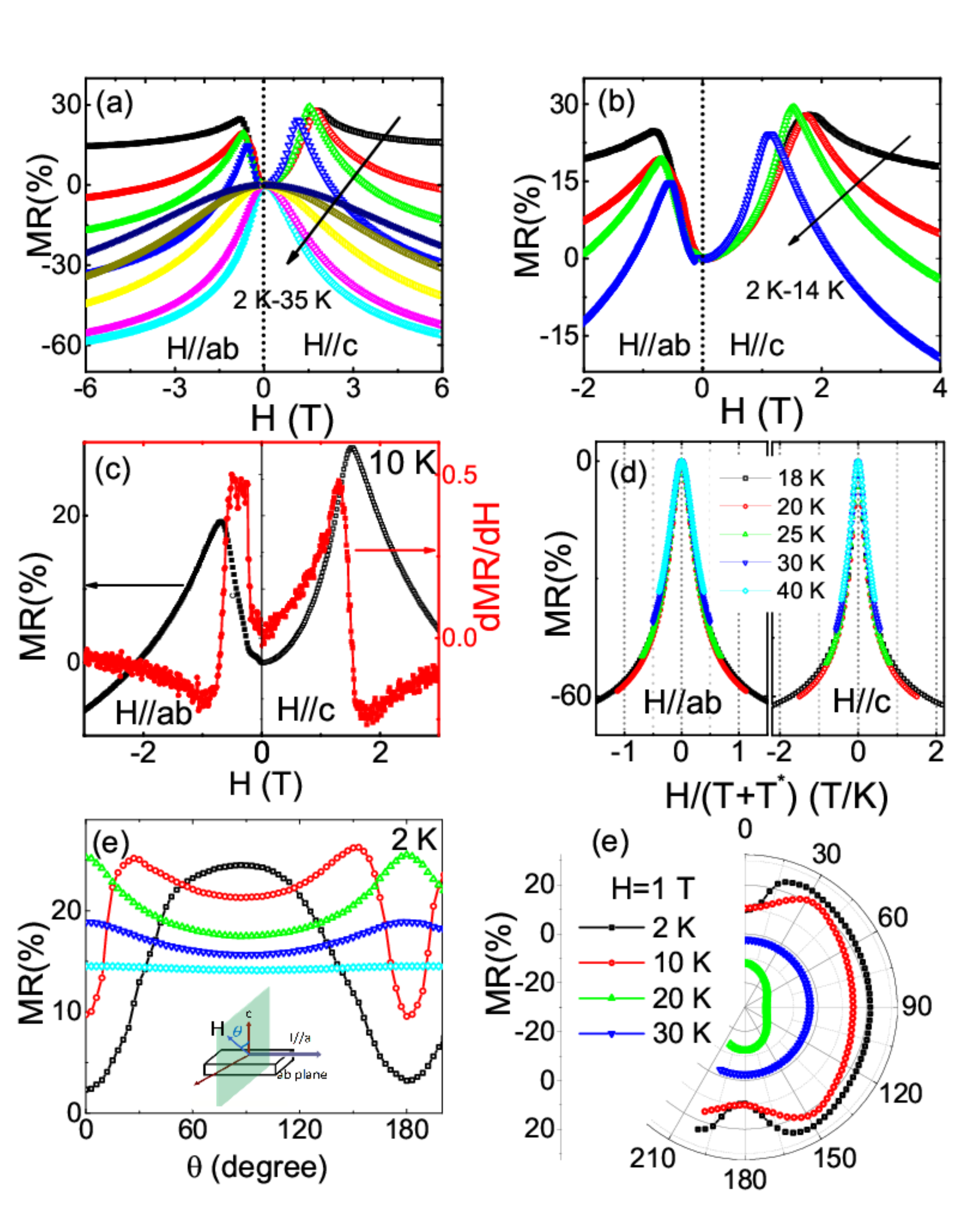}
  \caption{(a) MR for EuIn$_2$As$_2$ at 2 K, 6 K, 10 K,14 K, 16 K, 18 K, 20 K, and 30 K with $H$//$ab$ and $H$//$c$ respectively. (b) Zoomed MR for EuIn$_2$As$_2$ at 2 K, 6 K, 10 K,14 K with $H$//$ab$ and $H$//$c$ respectively.  (c) Zoomed MR and  dM/dH at 10 K with $H$//$ab$ and $H$//$c$ respectively. (d)  Scaled MR at 18 K, 20 K, 25 K, 30 K and 35 K with $H$//$ab$ and $H$//$c$ respectively.  (e) The AMR at 2 K with $H$=0.5 T, 1 T, 2 T, 3 T, and 6 T. Inset: Schematic configuration of the rotating sample. (f) Angular dependence of MR at 2 K , 10 K, 20 K, and 30 K with $H$=1 T}
  \label{fig:Fig5}
\end{figure}

In a magnetic system, the detailed magnetization could strongly affect the transport properties of a material such as the EuSn$_2$As$_2$ \cite{Xin Gui,Hang Li}. As shown in Fig. 5 (a) the MR for EuIn$_2$As$_2$ exhibits a monotone evolution  with both $H$//$ab$ and H//$c$ closely relating to the detailed magnet states. Below $T_N$ (18 K), the observed positive MR in the low field region increases with rising applied field to the maximum value around the saturation fields $H_{sa}$ and $H^\perp_{sa}$. Further increasing the magnetic field, the MR decreases and eventually become negative at higher fields. The peak in field dependent magnetoresitivity MR($H$) curve moves towards lower fields as well as reduces the intensity with increasing the temperature and finally disappears at $T_N$. Such a feature is characteristic of a metamagnetic phase transition \cite{Schlottmann} agreeable with our magnetization results (see Fig. 4). The difference of MR for $H$//$ab$ and $H$//$c$ is mainly due to the appearance of spin-flop states with $H$//$ab$ below $T_N$ shown in Fig. 5 (b). The kinked MR($H$)curves are observed for $H$//$ab$ near the spin-flop region while the MR varies smoothly with $H$ for $H$//$c$ below the saturation fields. The kink in $M$($H$) caused by the appearance of spin-flopped magnetic domains is more clear in the d$MR$/d$H$ at 10 K shown in Fig. 5 (c).  The various magnetic states mainly contribute to the anisotropic magneto-transport properties below $T_N$. Above $T_N$, the MR curves behave Kondo-like. In a Kondo-type AF system, the spin scattering by localized magnetic moments will be suppressed by aligning local moments giving rise to a negative MR\cite{Andrea M. Goforth}. By using the Bethe-Ansatz method\cite{Schlottmann}, the MR($H$) curves at various temperatures can be scaled on to a single curve with the relation:
 \begin{equation}
\mathrm{MR}(H)=f(\frac{H}{T+T^*})
\end{equation}
as shown in Fig. 5 (d). For $H$//$ab$ and $H$//$c$, the characteristic temperature $T^*$ is acquired to be -14.1 K and -14 K respectively. The  negative value indicates the presence of ferromagnetic correlations within the AF ground-state of EuIn$_2$As$_2$ \cite{Schlottmann}.

The angular dependent magnetoresistivity (AMR) is shown in Fig. 5 (e) and (f) with the measurement configuration shown in the inset of Fig. 5 (e). The current flows within the $ab$-plane and the magnetic field is applied and rotates within a plane perpendicular to the current. Magnetic field-induced changes in the AMR symmetry are observed. At 2 K , a two-fold symmetric MR is observed with a maximum value in the direction of $H$//$ab$ ($H$=0.5 T). Increasing the applied field, MR exhibits a maximum for the direction of $H$//c ($H$=2 T) due to the canted spins. Further increasing the field, the MR becomes isotropic as shown in Fig.5.(e). With increasing the temperature, the angular MR exhibits a similar symmetry evolving  with H=1 T and changes the value from positive to negative shown in Fig.5(g). Magnetic field-induced changes in the AMR have also been observed in other topological materials such as ZrTe$_5$ and EuTiO$_3$ \cite{G. Zheng,Kaveh Ahadi}. The applied magnetic field controls both the spin canting as well as Zeeman splitting leading to topological phase transitions such as the movement of Weyl points \cite{G. Zheng,Kaveh Ahadi}. For an AF topological insulator,  the detailed magnetization of such a system are key to the topological nature and controllable by various ways. As shown in this paper, EuIn$_2$As$_2$ exhibits various magnetic states tunable by temperature, magnetic field strength and direction suggesting the possibility of various topological states and transitions in this material.

\section{conclusion}
The electronic structure and transport properties are systemically studied for the ATI candidate EuIn$_2$As$_2$. Hole-type Fermi pockets are revealed and their linear energy dispersion near Fermi level agrees with the proposed topological state around the $\Gamma$-point. External magnetic fields applied  within the $ab$-plane lead to spin-flop transitions suggesting in-plane AF magnetic moments of the ground state. The  anisotropic magnetic phase diagrams are revealed and the temperature and field driven magnetic states lead to the systemic evolving of the transport properties in the low temperature region. These results reveal the detailed magnetization of EuIn$_2$As$_2$ and its potential to realize tunable axion states or other non trivial topological states.

Work at SYSU were supported by the Hundreds of Talents program of Sun Yat-Sen University, the Fundamental Research Funds for the Central Universities, NSFC-11904414, NSF of Guangdong under Contract No.2018A030313055, and Physical Research Platform (PRP) in School of Physics, SYSU. The ARPES experiments at HiSOR were performed with the approval of it Proposal Assessing Committee (Proposal Numbers 19AG004).

\end{document}